\documentclass[sigconf]{acmart}
\usepackage{hyperref}
\usepackage{amsmath}
\usepackage{algorithm}
\usepackage[noend]{algpseudocode}
\usepackage{adjustbox}
\usepackage{dblfloatfix}
\usepackage{booktabs}
\usepackage[T1]{fontenc}
\algnewcommand\algorithmicforeach{\textbf{for each}}
\algdef{S}[FOR]{ForEach}[1]{\algorithmicforeach\ #1\ \algorithmicdo}
\AtBeginDocument{%
  \providecommand\BibTeX{{%
    \normalfont B\kern-0.5em{\scshape i\kern-0.25em b}\kern-0.8em\TeX}}}

\setcopyright{none} 
\renewcommand\footnotetextcopyrightpermission[1]{} 
\begin{document}

\title[Generate FAIR Literature Surveys with Scholarly Knowledge Graphs]{Generate FAIR Literature Surveys\\with Scholarly Knowledge Graphs}

\author{Allard Oelen}
\orcid{0000-0001-9924-9153}
\affiliation{%
  \institution{L3S Research Center, Leibniz University Hannover, Germany}
}
\email{oelen@l3s.de}

\author{Mohamad Yaser Jaradeh}
\orcid{0000-0001-8777-2780}
\affiliation{%
  \institution{L3S Research Center, Leibniz University Hannover, Germany}
}
\email{jaradeh@l3s.de}


\author{Markus Stocker}
\orcid{0000-0001-5492-3212}
\affiliation{%
  \institution{TIB Leibniz Information Centre for Science and Technology, Germany}
}
\email{markus.stocker@tib.eu}

\author{S\"oren Auer}
\orcid{0000-0002-0698-2864}
\affiliation{%
  \institution{TIB Leibniz Information Centre for Science and Technology \& L3S Research Center, Germany}
}
\email{auer@tib.eu}

\renewcommand{\shortauthors}{Oelen et al.}

\begin{abstract}
Reviewing scientific literature is a cumbersome, time consuming but crucial activity in research. Leveraging a scholarly knowledge graph, we present a methodology and a system for comparing scholarly literature, in particular \emph{research contributions} describing the addressed problem, utilized materials, employed methods and yielded results. The system can be used by researchers to quickly get familiar with existing work in a specific research domain (e.g., a concrete research question or hypothesis). Additionally, it can be used to publish literature surveys following the FAIR Data Principles. The methodology to create a research contribution comparison consists of multiple tasks, specifically: (a) finding similar contributions, (b) aligning contribution descriptions, (c) visualizing and finally (d) publishing the comparison. The methodology is implemented within the Open Research Knowledge Graph (ORKG), a scholarly infrastructure that enables researchers to collaboratively describe, find and compare research contributions. We evaluate the implementation using data extracted from published review articles. The evaluation also addresses the FAIRness of comparisons published with the ORKG.
\end{abstract}


\maketitle

\section{Introduction}
When conducting scientific research, reviewing the existing literature is an essential activity~\cite{Seuring2005}. 
Familiarity with the state-of-the-art is required to effectively contribute to advancing it and do relevant research. Mainly because published scholarly knowledge is unstructured~\cite{Kuhn2016DecentralizedNanopublications}, it is currently very tedious to review existing literature. Relevant literature has to be found among hundreds and increasingly thousands of PDF articles. This activity is supported by library catalogs and online search engines, such as Scopus or Google Scholar~\cite{LasdaBergman2012}. Because the search is keyword based, typically large numbers of articles are returned by search engines. Researchers have to manually identify the relevant papers. Having identified the relevant papers, the relevant pieces of information need to be extracted in order to obtain an overview of the literature. Overall, these are manual and time consuming steps. We argue that a key issue is that the scholarly knowledge communicated in the literature does not meet the FAIR Data Principles~\cite{Wilkinson2016}. While PDF articles can be found and accessed (assuming Open Access or an institutional subscription), the scholarly literature is insufficiently interoperable and reusable, especially for machines. For units more granular than the PDF article, such as a specific result, findability and accessibility score low even for humans.

We present a methodology and its implementation integrated into the Open Research Knowledge Graph (ORKG)~\cite{jaradeh2019open} that can be used to generate and publish literature surveys in form of machine actionable, comparable descriptions of research contributions. Machine actionability of research contributions relates to the ability of machines to access and interpret the contribution data.  The benefits for researchers of such an infrastructure are (at least) two-fold. Firstly, it supports researchers in creating state-of-the-art overviews for specific research problems efficiently. Secondly, it supports researchers in publishing literature surveys that adhere to the FAIR principles, thus contributing substantially to reuse of state-of-the-art overviews and therein contained information, for both humans and machines. 

Literature reviews are articles that focus on analysing existing literature. Among other things, reviews can be used to gain understanding about a research problem or to identify further research directions~\cite{gall8c, Randolph2009}. Reviews can be used by authors to quickly obtain an overview of either emerging or mature research topics~\cite{Torraco2005}. Review papers are important for research fields to develop. When review papers are lacking, the development of a research field is weakened~\cite{Webster2002}. Compiling literature review papers is a complicated task~\cite{Wee2016} and is often more time consuming than performing original research~\cite{Webster2002}.  The structure of such articles often consists of tables that compare published research contributions. Although in the literature the terms ``literature review'' and ``literature survey'' are sometimes used interchangeably, we make the following distinction. We refer to the tables in review articles as \textit{literature surveys}. Together with a (textual) analysis and explanation, they form the \textit{literature review}. The state-of-the-art (SoTA) analysis is a special kind of literature review with the objective of comparing the latest and most relevant papers in a specific domain. 

We implement the presented methodology in the ORKG. The ORKG is a scholarly infrastructure designed to acquire, publish and process \emph{structured} scholarly knowledge published in the literature~\cite{jaradeh2019kcap}. ORKG is part of a larger research agenda aiming at machine actionable scholarly knowledge that understands the ability to more efficiently compare literature as a key feature. 

We tackle the following research questions: 

\begin{enumerate}
    \item How to generate literature surveys using scholarly knowledge graphs?
    \item How to ensure that published literature surveys comply with the FAIR principles?
     \item How to effectively specify and visualize literature surveys in a user interface?
\end{enumerate}

In support of the first research question, we present a methodology that describes the steps required to generate literature surveys. In support of the second research question, we describe how the FAIRness of the published literature review is ensured. Finally, in support of the third research question, we demonstrate how the methodology is implemented within the ORKG. 

The paper is structured as follows. Section~\ref{section:motivating-example} motivates the work. Section \ref{section:related-literature} reviews related work. Section~\ref{section:system-design} presents the system design, the underlying methodology and its implementation. Section~\ref{section:data-collection} explains how the knowledge graph is populated with data. Section~\ref{section:evaluation} presents the evaluation of the system, specifically system FAIRness and performance. Finally, Section~\ref{section:discussion} discusses the presented and future work.

\section{Use cases}
\label{section:motivating-example}

We motivate our work by means of two use cases that underscore the usefulness of a literature survey generation system. In the first use case, a researcher wants to obtain an overview on state-of-the-art research addressing a specific problem. The second use case describes how a researcher can publish a FAIR compliant literature review with the ORKG. 

\paragraph{Familiarize with the state-of-the-art}
A state-of-the-art (SoTA) analysis reviews new and emerging research. They are useful for multiple reasons. Firstly, they provide a broad overview of a research problem and support understanding. Secondly, they juxtapose different approaches for a problem. Thirdly, they can support claims on why certain research is relevant by giving an overview of the breadth of research addressing a problem. 
The proposed approach enables automated generation of surveys to quickly obtain an overview of state-of-the-art research as well as sharing of surveys for others to reuse.


\paragraph{Publishing of literature reviews}
\label{section:motivating-example:publish-reviews}
Literature reviews typically consist of multiple (survey) tables in which different approaches from original papers are compared based on a set of properties. These tables can be seen as the main contribution and most informative part of the review paper, since the tables juxtapose and compare existing work. Comparison tables are published in review papers as static content in PDF documents. This presentational format is generated from datasets that typically contain more (structured) information than what is presented in the published table. However, the additional information is not published. It is ``dark data'' which is not stored or indexed and likely lost over time~\cite{Heidorn2008}. Furthermore, published tables are not machine actionable. Their overall low FAIRness hinders reusability of the published content. With the presented service, it is possible to publish a literature survey with high FAIRness, i.e. that is compliant with the FAIR principles to a high degree. Section~\ref{section:related-literature} discusses this aspect in more details.

\paragraph{Summary of weaknesses of the current approach to literature review}

The weaknesses of the current approach to literature review can be summarized as follows:

\begin{itemize}
    \item {\it Static} -- reviews are static, since once published as PDF they are rarely updated and there are no possibilities or incentives for creating new or updated reviews for considerable time.
    \item {\it Lack of machine assistance} -- machine assistance is hardly possible, since the PDF representation of reviews is only human readable and relevant raw data is mostly not published along with the review.
    \item {\it Delay} -- reviews are produced and published with significant delay (often years) after original research work was done.
    \item {\it Coverage} -- due to the amount of work required, reviews are often only performed for relatively popular research topics and are stale or missing for less popular topics.
    \item {\it Lacking collaboration} -- collaboration on reviews is not possible and reviews currently represent only the viewpoint of the few authors not the community.
    \item {\it Missing overarching systematic semantic representation} -- the overlap between different reviews and related work sections in individual original research papers is not explicit and cannot be exploited.
\end{itemize}

We deem that these weaknesses of the current approach to scholarly literature review and synthesis significantly hinder scientific progress.

\section{Related work}
\label{section:related-literature}
The task of comparing research contributions can be reviewed in light of the more general task of comparing resources (or entities) in a knowledge graph. While this is a well-known task in multiple domains (for instance in e-commerce systems~\cite{ziemba2017online}), not much work has focused on comparison in knowledge graphs, specifically. One of the few works with this focus is by \citeauthor{Petrova2017}~\cite{Petrova2017} who created a framework for comparing entities in RDF graphs using SPARQL queries. In order to compare contributions, they first have to be found. Finding is an information retrieval problem. As a well-known technique, TF-IDF~\cite{Medina2015} can be used for this task. More sophisticated techniques can be used to determine the structural similarity between graphs (e.g., \cite{Maillot2019}) and matching semantically similar predicates. This relates to dataset interlinking~\cite{Araujo2011} or more generally ontology alignment~\cite{Shvaiko2013}. For property alignment, techniques of interest include edit distance (e.g., Jaro-Winkler~\cite{Winkler1990} or Levenshtein~\cite{levenshtein1966binary}) and vector distance. \citeauthor{Gromann2019}~\cite{Gromann2019} found that fastText~\cite{Bojanowski2017} performs best for ontology alignment. 

In light of the FAIR Data Principles~\cite{Wilkinson2016}, scholarly data should be Findable, Accessible, Interoperable and Reusable both for humans and machines. Due to the publication format, literature survey tables published in scholarly articles weakly adhere to the FAIR guidelines, particularly so for machines. Scholarly data should be considered first-class objects~\cite{Starr2015}, including data used to create literature surveys. \citeauthor{Rodriguez-Iglesias2016}~\cite{Rodriguez-Iglesias2016} describe the difficulties of making data FAIR within the plant sciences. They argue that it is more complicated than reformatting data. On the other hand they suggest that most FAIR principles can be implemented relatively easily by using off-the-shelf technologies. \citeauthor{Boeckhout2018}~\cite{Boeckhout2018} argue that the FAIR principles alone are not sufficient to lead to responsible data sharing. More applied principles are needed to ensure better scholarly data. This claim is supported by the findings of \citeauthor{Mons2017}~\cite{Mons2017} who suggest that there are very diverse interpretations of the guidelines. In their work, they try to clarify what is FAIR and what is not. 

An efficient literature comparison relies on scholarly knowledge being represented in a structured way. There is substantial related work on representing scholarly knowledge in structured form~\cite{IniestaCorcho:SePublica2014}. Building on the work of numerous philosophers of science, \citeauthor{Hars2001}~\cite{Hars2001} proposed a comprehensive scientific knowledge model that includes concepts such as theory, methodology and statement. More recently, ontologies were engineered to describe different aspects of the scholarly communication process. Semantic Publishing and Referencing (SPAR)\footnote{http://purl.org/spar/\{cito,c4o,fabio,biro,pro,pso,pwo,doco,deo\}} is a collection of ontologies that can be used to describe scholarly publishing and referencing of documents~\cite{Peroni2012,Gangemi2017,peroni2018spar,Constantin2016}. \citeauthor{IniestaCorcho:SePublica2014}~\cite{IniestaCorcho:SePublica2014} reviewed the state-of-the-art ontologies to describe scholarly articles. \citeauthor{Sateli2015SemanticRO}~\cite{Sateli2015SemanticRO} use some of these scholarly ontologies to add semantic representations of scholarly articles to the Linked Open Data cloud. A literature survey comparing scholarly ontologies is available via the ORKG.\footnote{\url{https://www.orkg.org/orkg/comparison/R8342}} Most of these ontologies are designed to capture metadata about and structure of scholarly articles, not the content communicated in articles. 
Another literature survey is created to compare approaches for semantically representing scholarly communication.\footnote{\url{https://www.orkg.org/orkg/comparison/R8364}}


An initial attempt for semantifying review articles was done in~\cite{DBLP:conf/ercimdl/FathallaVA017}. The work comprises a relatively rigid ontology for describing contributions (mainly centered around research problems, approaches, implementations and evaluations) and a prototypical implementation using Semantic MediaWiki. We relax this constraint, since we are not limited by a rigid ontology schema but rather allow arbitrary domain-specific semantic structures for research contributions. The work by \citeauthor{DBLP:conf/ercimdl/VahdatiFALV19}~\cite{DBLP:conf/ercimdl/VahdatiFALV19} focuses on semantic article representations for generating literature overviews. Their method is to use crowdsourcing to generate the overviews. \citeauthor{Kohl2018}~\cite{Kohl2018} present CADIMA, a system that supports systematic literature reviews. The tool supports the formal process of performing a literature review but does, for example, not publish data in machine actionable form for reuse.

\begin{figure}[t]
    \centering
    \includegraphics[width=\columnwidth]{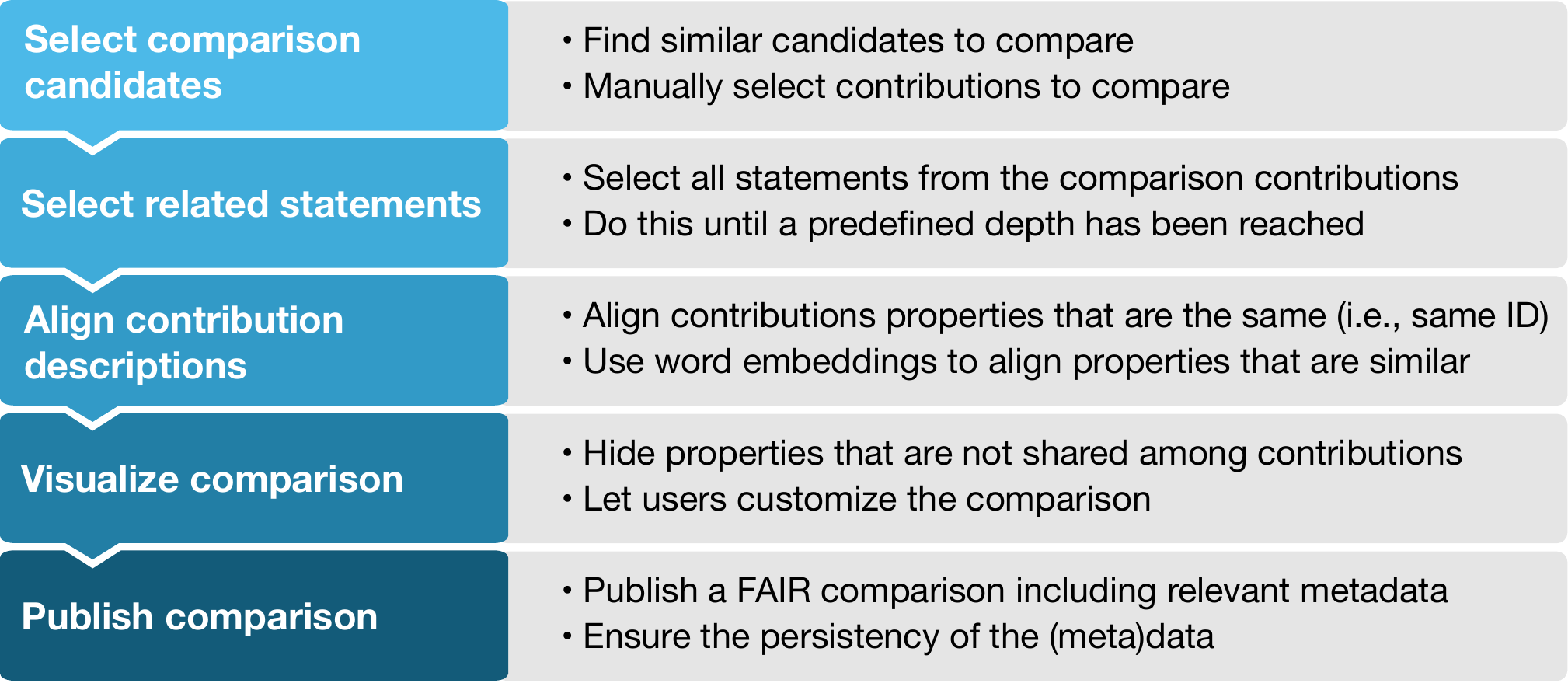}
    \caption{Research contribution comparison methodology.}
    \label{fig:workflow-comparison}
\end{figure}

\section{System design}
\label{section:system-design}

We now present the system design of the literature comparison service. It consists of a methodology that describes how to perform a comparison of research contributions. An early version of this methodology has been presented at the 3rd SciKnow workshop~\cite{Oelen:Sciknow19}. The methodology consists of five steps: 1) finding comparison candidates, 2) selecting related statements, 3) aligning contribution descriptions, 4) visualizing comparisons and 5) publishing FAIR comparisons. The methodology is depicted in Figure~\ref{fig:workflow-comparison}. First, we discuss the data structure of the ORKG, which forms the foundation of the comparison. Then, each step of the methodology is described in more detail. Finally, we discuss the implementation. 

\subsection{ORKG ontology}
\label{section:ontology}
In ORKG, each paper is typed as \textit{paper} class. A paper consists of at least one \textit{research contribution}, which addresses at least one \textit{research problem}. Research contributions consist of \textit{contribution data} that describe the contribution. For instance, a paper in Computer Science might have descriptions for materials, methods, implementation and results as contribution data. These predefined core concepts can be easily extended with domain specific research problems, methods, etc. in ORKG curation using crowdsourcing or other curation approaches. The underlying data structure uses the notion of statements. Statements are triples that consist of a subject, a predicate (also called a property) and an object.  The granularity of a comparison is at the research contribution, meaning that contributions are compared rather than papers. For simplicity, we use the terms ``paper comparison'' and ``contribution comparison'' interchangeably. Because a comparison happens on contribution level, it is possible to compare specific elements of a paper instead of the complete paper. The benefit of this is that a comparison does not contain data from irrelevant contributions. The ORKG OWL ontology is available online.\footnote{\url{https://gitlab.com/TIBHannover/orkg/orkg-ontology}}  


\subsection{Select comparison candidates}
\label{section:workflow-comparison-candidates}
To perform a comparison, a starting contribution is needed. This contribution is called \textit{main contribution} and is always manually selected by a user. The main contribution is compared against other \textit{comparison contributions}. There are two different approaches for selecting the comparison contributions. The first approach automatically selects comparison contributions based on similarity. The second approach lets users manually select contributions. 

\subsubsection{Find similar contributions}
Comparing contributions makes only sense when contributions can sensibly be compared. For example, it does not make (much) sense to compare a biology paper to a history paper. 
We thus argue that it makes only sense to compare contributions that are similar. More specifically, contributions that share the same (or a similar set of) properties are good comparison candidates. For instance, a paper about question answering has the property \textit{orkg:disambiguationTask}\footnote{\textit{orkg:} denotes the ontology of the ORKG system described in Section~\ref{section:ontology}} and another paper is using the same property to describe what disambiguation tasks are performed. Since they share the same property it makes them likely candidates for comparison. Finding similar contributions is therefore based on finding contributions that share the same or similar informative description properties. To achieve this, each comparison contribution is converted into a string by concatenating all properties of the contribution. 
TF-IDF~\cite{Medina2015} is used to query these strings with the string of the main contribution as query. The search returns the most similar contributions by weighting the most informative properties higher due to TF-IDF. The top-k contributions are selected and form a set of contributions that are used in the next step.


\begin{figure}[t]
    \centering
    \includegraphics[width=0.9\columnwidth]{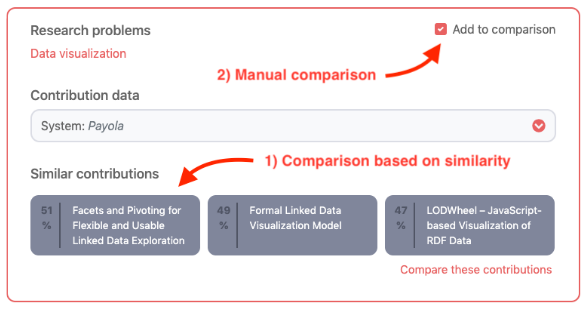}
    \caption{Implementation of the first step of the methodology: the selection of comparison candidates. Showing both the similarity-based and the manual selection approaches.}
    \label{fig:comparison-candidates}
\end{figure}

Figure~\ref{fig:comparison-candidates} displays how the similar contribution selection is implemented. As depicted, three similar contributions are suggested to the user (with the corresponding similarity percentage being displayed next to paper title). These suggested contributions can be directly compared.

\subsubsection{Manual selection}
\begin{figure}[t]
    \centering
    \includegraphics[width=.45\columnwidth]{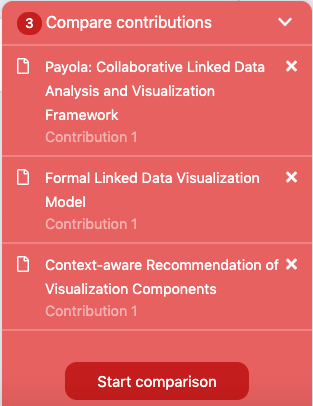}
    \caption{Box showing the manually selected contributions.}
    \label{fig:comparison-box}
\end{figure}

There are scenarios where comparison based on similarity computation is not suitable or desired. For example, a researcher wants to compare a specific set of implementations to see which performs best. 
Therefore, the manual selection method is implemented in a similar fashion to an e-commerce shopping cart. When the ``Add to comparison'' checkbox is checked, a box appears listing the selected contributions (Figure~\ref{fig:comparison-box}). 

\subsection{Select related statements}
\label{section:workflow-related-statements}
This step selects the statements from the graph related to the set of contributions selected in the previous step. Statements are selected transitively to match contributions in subject or object position. This search is performed until a predefined maximum transitive depth $\delta$ has been reached. The intuition is that the deeper a property is nested the less likely is its relevance for the comparison. The process of selecting statements is repeated until depth $\delta=5$ is reached. This number is chosen empirically to include statements that are not directly related to the contribution, but to exclude statements that are less relevant because they are nested too deep. 


\subsection{Align contribution descriptions}
As described in the first step, comparisons are built using shared or similar properties of contributions. In case the same property has been used between contributions, these properties are grouped and form one \textit{comparison row}. However, often different properties are used to describe the same concept. This occurs for various reasons. The most obvious reason is when two different ontologies are used to describe the same property. For example, for describing the population of a city, DBpedia uses \textit{dbo:populationTotal} while WikiData uses \textit{WikiData:population} (actually the property identifier is P1082; for the purpose here we use the label). When comparing contributions, these properties should be considered as equivalent. Especially for community-created knowledge graphs, differently identified properties likely exist that are, in fact, equivalent.

To overcome this problem, we use pre-trained fastText~\cite{Bojanowski2017} word embeddings to determine the similarity of properties. If the similarity is higher than a predetermined threshold $\tau$, the properties are considered equivalent and are grouped. This happens when the similarity threshold $\tau \geq 0.9$ (also empirically determined). In the end, each group of properties will be visualized as one row in the comparison table. The result of this step is a list of statements for each contribution, where similar properties are grouped. Based on this similarity matrix $\gamma$ is generated 

\begin{equation}
    \gamma_{p_{i}} = \left [ cos(\overrightarrow{p_i}, \overrightarrow{p_j}) \right ]
    \label{eq:similarity_matrix}
\end{equation}

with $cos(.)$ as the cosine similarity of vector embeddings for property pairs $(p_i, p_j) \in \mathcal{P}$, whereby $\mathcal{P}$ is the set of all contributions.

Furthermore, we create a mask matrix $\Phi$ that selects properties of contributions $c_i \in \mathcal {C}$, whereby $\mathcal{C}$ is the set of contributions to be compared. Formally,

\begin{equation}
    \Phi_{i,j} = \begin{cases}
1 & \text{ if } p_{j} \in c_i \\ 
0 & \text{ otherwise }  
\end{cases}
    \label{eq:mask_matrix}
\end{equation}

Next, for each selected property $p$ we create the matrix $\varphi$ that slices $\Phi$ to include only similar properties. Formally,

\begin{equation}
\varphi_{i,j} =(\Phi_{i,j})_{\substack{c_i \in \mathcal{C}\\ p_j \in sim(p) }}
\label{eq:slice_mask}
\end{equation}

where $sim(p)$ is the set of properties with similarity values $\gamma[p] \geq \tau$ with property $p$. Finally, $\varphi$ is used to efficiently compute the common set of properties~\cite{jaradeh2019kcap}. This process is displayed in Algorithm~\ref{alg:alining-properties}.

\begin{algorithm}
\caption{Align contribution descriptions}\label{alg:alining-properties}
\begin{algorithmic}[1]
\Procedure{AlignProperties}{properties, threshold}
\ForEach {property $p_1 \in properties$}
    \ForEach {property $p_2 \in properties$}
        \State $similarity \gets$ cos(Embb($p_1$), Embb($p_2$))
        \If {$similarity > threshold$}
        \State $similarProps \gets similarProps \cup \{ p_1, p_2 \}$
        \EndIf
    \EndFor
\EndFor
\Return $similarProps$
\EndProcedure
\end{algorithmic}
\end{algorithm}

\subsection{Visualize comparison}
The next step of the workflow is to visualize the comparison and present the data in a human understandable format. Tabular format is often appropriate for visualizing comparisons since tables provide a good overview of data. Another aspect of the visualization is determining which properties should be displayed and which ones should be hidden. A property is displayed when it is shared among a predetermined amount $\alpha$ of contributions, where $\alpha$ mainly depends on comparison use and can be determined based on the total amount of contributions in the comparison. By default, only properties that are common to at least two contributions ($\alpha \geq 2$) are displayed. 

Another aspect of comparison visualization is the possibility to customize the resulting table. This is needed because of the similarity-based matching of properties and the use of predetermined thresholds. For example, users should be able to enable or disable properties. They should also get feedback on property provenance (i.e., the property's path in the graph). Ultimately, this contributes to a better user experience, with the possibility to manually correct mistakes made by the system. 

\begin{figure}[t]
    \centering
    \includegraphics[width=0.95\columnwidth]{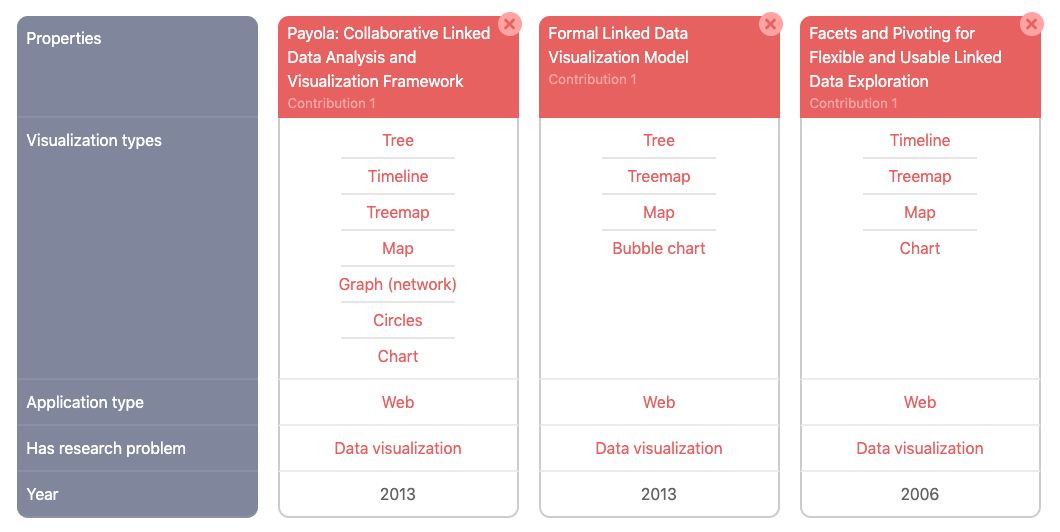}
    \caption{Comparison of research contributions related to visualization tools.}
    \label{fig:comparison-table}
\end{figure}

Figure~\ref{fig:comparison-table} displays a comparison for research contributions related to visualization tools published in the literature. In this example, four properties are displayed. Literals are displayed as plain text while resources are displayed as links. When a resource link is selected, a popup is displayed showing the statements related to this resource. The UI implements some additional features that are particularly useful to compare research contributions. 

\paragraph{Customization}
Users can customize comparisons including transposing the table as well as hiding and rearranging the properties. Especially the option to hide properties is helpful when contributions with many statements are compared. Only properties considered relevant to the user can be selected to display. Customizing the comparison table can be useful before exporting or sharing the comparison. 

\paragraph{Sharing and persistence}
Comparisons can be shared using a persistent link. Especially when sharing the comparison for research purposes, it is important to refer to the original comparison. Since contribution descriptions may change over time comparisons may also change. To support persistency, the whole state of the comparison is stored in a document-oriented database and retrieved when the permalink is invoked. 

\paragraph{Export}
It is possible to export comparisons in different output formats such as PDF, CSV, RDF and \LaTeX. The \LaTeX~export is useful for direct integration in research papers. Together with the \LaTeX~table, a BibTeX file containing the bibliographic information of the papers used in the comparison is also generated. Also, a persistent link referring back to the comparison in ORKG is showed as table footnote.

\subsection{Publish comparison}
\begin{figure}[t]
    \centering
    \includegraphics[width=.85\columnwidth]{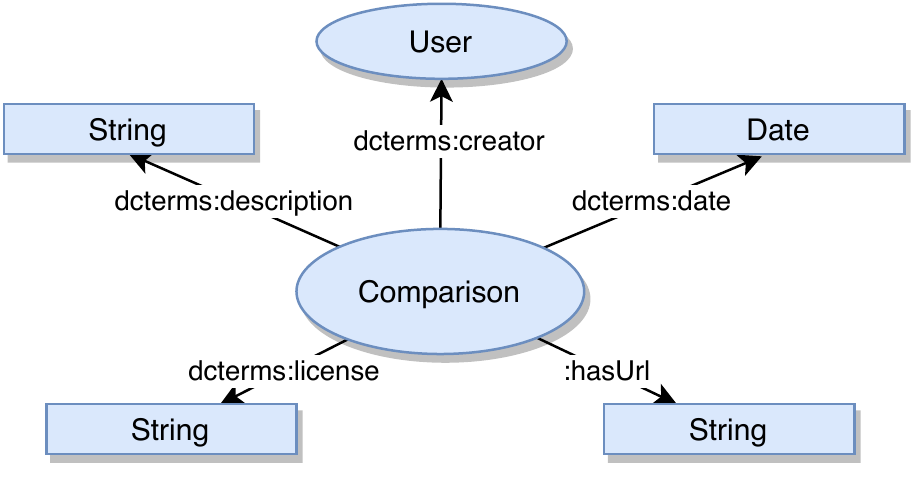}
    \caption{The graph structure of the metadata for a published comparison. The \textit{dcterms:} prefix denotes the Dublin Core Metadata Terms ontology.}
    \label{fig:RDF-comparison}
\end{figure}

Visualized and customized comparison tables can be stored. Storing tables is part of the publishing process and therefore only needed when a generated table is going to be used in a paper. In order to regenerate the table the whole state of the comparison should be saved. The knowledge graph from which the comparison was generated changes over time and thus storing just the URIs of the respective papers would not suffice. While saving a comparison, the user can provide additional metadata to ensure findability, an aspect of the FAIR principles. Metadata include a comparison title, which would normally consist of a one sentence description of the comparison. Additionally, a longer textual description can be provided. This metadata is extended with machine generated data, such as the creation date and the creator of the comparison. The metadata is stored in the knowledge graph to support easy access and interoperability. In Figure~\ref{fig:RDF-comparison}, the structure of the metadata is displayed using the Dublin Core Metadata Terms\footnote{\url{https://dublincore.org/specifications/dublin-core/dcmi-terms}}. The comparison data itself is stored in a document-oriented database. An RDF export of both the metadata and the comparison data can be generated. The comparison data is modeled with the RDF Data Cube Vocabulary\footnote{\url{https://www.w3.org/TR/vocab-data-cube}}. A unique identifier is attached when the comparison is saved. This ID is used when the comparison is shared or when it is referenced in a paper. The literature comparison can also be performed without publishing. Although the workflow and the steps to create a comparison stay the same, the goal is different. Instead of creating a comparison that will be published and referenced in a paper, the comparison will be used by the researcher herself.

\subsection{Technical details}
The user interface of the comparison feature is seamlessly integrated with the ORKG front end, which is written in JavaScript and is publicly available\footnote{\url{https://gitlab.com/TIBHannover/orkg/orkg-frontend}}. The back end of the comparison feature is a service separate from the ORKG back end written in Python and also available Open Source\footnote{\url{https://gitlab.com/TIBHannover/orkg/orkg-similarity}}. The comparison back end is responsible for step two and three of the comparison methodology. The input in step two is the set of contribution IDs. The API selects the related statements and aligns the properties and returns the data needed to visualize the comparison. This data includes the list of papers, list of all properties and the values per property. 

\begin{table*}[t]
\caption{List of imported survey tables in the ORKG. The paper and table reference can be used to identify the original table. }
\label{table:imported-review-tables}
\begin{adjustbox}{scale=.90}
\begin{tabular}{l|l|l|r|l|l}
\toprule
\textbf{Paper reference} & \textbf{Table reference} & \textbf{Research problem} & \textbf{Papers} & \textbf{ORKG representation} & \textbf{Information loss} 

\\ \midrule
\citeauthor{Bikakis2016}~\cite{Bikakis2016} & 
Table 1 & Generic visualizations & 
11 &
\url{https://orkg.org/orkg/c/pdLJDk} & 
No 
\\ 

\citeauthor{Bikakis2016}~\cite{Bikakis2016} & 
Table 2 & Graph visualizations & 
21 &
\url{https://orkg.org/orkg/c/Rx476Z} & 
No
\\ 

\citeauthor{Diefenbach2018a}~\cite{Diefenbach2018a} & 
Table 2 & 
Question answering evaluations & 
33 &
\url{https://orkg.org/orkg/c/gaVisD} & 
No
\\ 

\citeauthor{Diefenbach2018a}~\cite{Diefenbach2018a} & 
Table 3,4,5,6 & 
Question answering systems & 
26 &
\url{https://orkg.org/orkg/c/IuEWl2} & 
No
\\ 

\citeauthor{Hussain2017a}~\cite{Hussain2017a} & 
Table 4 &
Author name disambiguation &
5 &
\url{https://orkg.org/orkg/c/vDxKdr} & 
No
\\

\citeauthor{Hussain2017a}~\cite{Hussain2017a} & 
Table 5 &
Author name disambiguation &
6 &
\url{https://orkg.org/orkg/c/XXg8Wg} & 
No
\\

\citeauthor{Hussain2017a}~\cite{Hussain2017a} & 
Table 6 &
Author name disambiguation &
9 &
\url{https://orkg.org/orkg/c/9rOwPV} & 
No
\\

\citeauthor{Hussain2017a}~\cite{Hussain2017a} & 
Table 7 &
Author name disambiguation &
6 &
\url{https://orkg.org/orkg/c/mB7kIK} & 
No
\\ 

\citeauthor{Naidu2018}~\cite{Naidu2018} & 
Table 4 &
Text summarization &
52 &
\url{https://orkg.org/orkg/c/OUqYB9} & 
No
\\ 
\bottomrule
\end{tabular}
\end{adjustbox}
\end{table*}

\section{Data collection}
\label{section:data-collection}

In order to generate useful literature reviews it is crucial for the knowledge graph to contain sufficient and relevant papers. Populating the knowledge graph with high quality paper descriptions it not straightforward. Structured descriptions of papers should be created in such a way that it is possible to compare papers based on shared properties. Both published papers and papers that will be published in the future should be added to the ORKG, retrospectively or prospectively. Although a comprehensive description on how to populate the ORKG is out-of-scope here, we now briefly describe how we envision populating the ORKG in a manner that would facilitate comparing contributions.

Prospectively, authors can become part of generating structured descriptions of their papers. This should be done in a crowdsourced manner and can become part of the paper submission process. Input templates that collect relevant properties can be used to ensure structured and comparable paper descriptions. Retrospectively, automated (machine learning) methods can be helpful ensure scalability of the process of adding a paper.

\subsection{Leverage legacy review paper tables}
To populate the ORKG with comparable paper descriptions, we leverage the data published in review papers. Review papers consist of high quality, curated and often structured data that is collected from a set of papers that address the same (or a similar) research problem. Hence, using reviews to populate a scholarly knowledge graph is a relatively straightforward approach to obtain high quality structured paper descriptions. We now present a methodology to convert survey paper data into a knowledge graph structure. The steps are as follows:

\begin{enumerate}
    \item \textbf{Survey paper selection.} The first step is the selection of survey papers that are suitable for building a knowledge graph. Firstly, the survey should compare peer-reviewed scientific articles. For instance, a comparison of different systems without a reference to peer-reviewed work is not suitable for the scholarly knowledge graph. Secondly, the review should compare the papers' content in a structured way and should not merely list work in a field. Especially reviews that present their results and literature comparisons in tabular format are suitable. The result of this step is a list of papers that will be added to the ORKG. 
    \item \textbf{Table selection.} Given the selected survey papers, tables have to be selected. Some surveys contain only one table while in others multiple tables are presented. In some cases a collection of tables can be joined into one larger table. 
    \item \textbf{Data modeling.} Given the selected tables, a suitable graph structure has to be determined. The data structure has to be modeled. For instance, when implemented systems are compared, a suitable structure could be: \texttt{[has implementation] -> System name}. The referenced system can be described with a list of properties to be compared. Additionally, a research problem has to be defined, which is typically the same for all papers that are part of the table. 
    \item \textbf{Metadata collection.} Next, the metadata for the papers that are referenced in the survey table is collected. In case a referenced paper has a DOI\footnote{Digital Object Identifier}, the metadata can be automatically retrieved via a lookup service (e.g. Crossref\footnote{\url{https://www.crossref.org}}). Otherwise, at least the title, authors and publication date have to be collected.
    \item \textbf{Data ingestion.} Finally, the paper data is ingestion into the knowledge graph. The paper data consists of both the paper's metadata and the extracted data from the comparison table. 
    This does not result in a single description of the survey paper. Each paper referenced in the survey table is ingested individually. In order to speed up the process of adding papers, we developed a Python package\footnote{\url{https://gitlab.com/TIBHannover/orkg/orkg-pypi}} that has a function to add a paper to the knowledge graph. 
\end{enumerate}
This methodology has been used to populate the ORKG with comparable paper data. The data is used to evaluate the presented literature review tool. The imported paper data is not only useful for the evaluation, but does also provide significant value to the ORKG itself.

\begin{figure}[t]
    \centering
    \includegraphics[width=0.8\columnwidth]{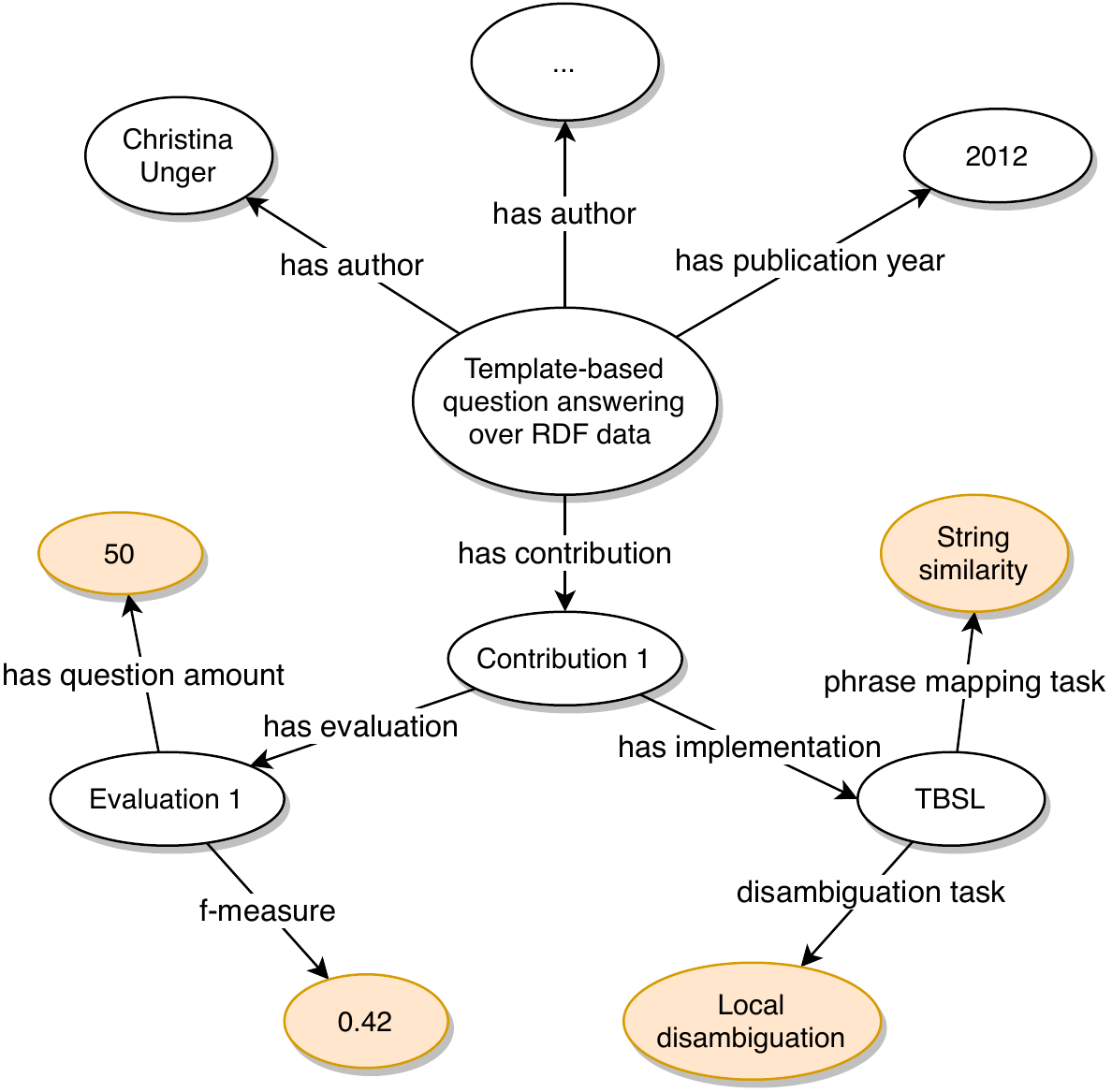}
    \caption{Partial graph structure of an imported paper. Orange colored resources indicate potentially interesting values for a paper comparison.}
    \label{fig:graph-example-import}
\end{figure}

In total, four review papers were selected for importing into the ORKG. The Python script for importing the table data is available online.\footnote{\url{https://gitlab.com/TIBHannover/orkg/orkg-papers}} From those papers, 12 different tables were imported. Together, 169 papers were reviewed in those four survey papers. This resulted in a total amount of $3\,750$ statements being added to the knowledge graph. Table~\ref{table:imported-review-tables} lists the imported review papers and tables. The survey papers address different research problems. Figure~\ref{fig:graph-example-import} depicts an excerpt of the resulting graph for one particular paper. A set of comparison tables made with the imported data is available online.\footnote{\url{https://orkg.org/orkg/featured-comparisons}} This list includes some alternative comparison tables that were generated with the same data. 

\section{Evaluation}
\label{section:evaluation}

In this section, we present an evaluation of multiple aspects of the presented comparison methodology and implementation. Firstly, we evaluate information representation. Then, we evaluate the FAIRness of published reviews. Finally, we present a performance evaluation that tests the scalability.

\subsection{Information representation}
This part of the evaluation focuses on the aspect of information representation. We use the data from the imported review papers, as described in Section~\ref{section:data-collection}. In order to build and publish useful and correct literature reviews, at a minimum our service should display the same information that was originally presented in the review tables. This means that there should not be information loss when review tables are published using our service. If there is no information loss, it means our service can be used as an alternative to the current way of publishing review tables. Apart from generating the same table, the added value comes from the ability to aggregate new (tabular) views using the same data as well as the increased FAIRness of the data published via our service. For each of the imported review tables, listed in Table~\ref{table:imported-review-tables}, we can evaluate whether the same table can be generated with our service. For this, we have compared the table from the review paper to the table generated by the ORKG comparison service. A collection of 169 paper with 9 distinct literature views/tables are part of this evaluation. These tables can be viewed online, the links are listed in the ``ORKG representation'' column. The results of this evaluation are displayed in the same table, in column ``Information loss''. As the results show, using our service it is possible to recreate the same tabular views as originally published in the review papers.

\subsection{FAIR data evaluation}
As described before, with the presented service it is possible to publish a generated comparison that adheres with the FAIR principles. Because the service leverages a knowledge graph to generate and save comparisons, complying with the FAIR principles is more obvious for the ORKG comparison service than for tables in published PDF articles. In order to evaluate the FAIRness of a published comparison, we evaluate each of the four FAIR principles in detail. \citeauthor{Wilkinson2016}~\cite{Wilkinson2016} described each principle by assigning sub-principles.\footnote{For a more detailed definition of the FAIR principles, see: \url{https://go-fair.org/fair-principles}} We discuss the relevant sub-principles and explain how they are met. We use the term (meta)data to refer to both the actual comparison data (i.e., the data that is used to create the comparison table) and the associated metadata (e.g., the title, description and creator of a comparison). Table \ref{table:fair-principles-compliance} presents an overview for the evaluation of the FAIR principles.

\paragraph{Findable.} To make data findable for both humans and machines (i.e., agents), a unique and persistent identifier should be attached to the data (F1). Additionally, metadata should describe the data (F2). In the metadata, the unique identifier of the data should be mentioned (F3). Also a search interface should be available to find the data (F4). To ensure the findability of comparisons, users can title and describe them. Furthermore, machine generated metadata is attached to a comparison (e.g., the number of papers and the creation date). A unique identifier is generated and attached to the data and included in the metadata. Finally, the ORKG search interface allows users to search the whole graph and has a dedicated filter to specifically find comparisons. Additionally, comparisons can be indexed and found by third-party search engines (such as Google or Bing).
    
\paragraph{Accessible.} Having found data, agents need to know how to access it. This principle is primarily about using accessible standardised communication protocols (A1). Additionally, metadata should be available even when the data is not (A2). The metadata is part of the knowledge graph, which can be accessed via the HTTP protocol. The data can be accessed without authentication. To support A2, the metadata and the actual comparison data are stored separately. Therefore, it is possible to access only metadata when the original data is not available anymore (for example when data is retracted by the author). 

\paragraph{Interoperable.} To ensure the interoperability of data, it should use a formal language for knowledge representation (I1) and should use vocabularies that are FAIR (I2). Finally, references or links to other (meta)data should be made (I3). As argued before, thanks to highly structured data and the integration of shared vocabularies, interoperability is an inherent feature of knowledge graphs. Data is (partially) described using the ORKG core ontology and other ontologies we use to canonicalize the representation of relevant information content types. Links to other data are present in the knowledge graph. For example, if a comparison uses the ``Web'' resource to specify the domain of an application, this resource is generic, can be shared among paper descriptions and comparisons, and can be described in more detail, independently of a particular comparison.

\paragraph{Reusable.} Finally, data should be reuseable. This can be accomplished by adding relevant (meta)data (R1). Required are an accessible data license (R1.1) and detailed provenance (R1.2) data. Finally, (meta)data should use community standards to describe data (R1.3). It is possible to add additional metadata to a comparison, e.g. metadata about the scope of the comparison, which could be a reference to the paper in which the comparison is being used. The metadata is complemented with the metadata that is already part of the Findability principle, e.g. provenance data about the creator of the comparison. The data license of the graph data is CC BY-SA\footnote{\url{https://creativecommons.org/licenses/by-sa/2.0}} (Attribution-ShareAlike), which allows reuse of the data. There is currently no community standard to describe the comparison data. However, standard ontologies are used to describe metadata (e.g., Dublin Core).

\begin{table}[t]
\centering
\caption{Overview of FAIR principles compliance.}
\label{table:fair-principles-compliance}
\begin{adjustbox}{scale=.85}
\begin{tabular}{l|c|p{67mm}}
\toprule
\textbf{Principle} & \textbf{Level} & \textbf{Explanation} \\ \midrule
\multicolumn{3}{l}{\textit{Findable}} \\ \midrule
F1 & 3 & Unique IDs exist, DOI assignment for future work \\ 
F2 & 2 & Machine and user generated metadata is attached \\ 
F3 & 1 & Properties used to link data to metadata \\ 
F4 & 1 & Comparisons are findable via a search interface \\ \midrule
\multicolumn{3}{l}{\textit{Accessible}} \\ \midrule
A1 & 2 & Data is accessed over HTTP (via REST or a user interface), requires user effort to integrate the ORKG API specification \\ 
A1.1 & 1 & The protocol is free and widely used \\ 
A1.2 & 1 & No authentication is required to access the data \\ 
A2 & 1 & Metadata is stored in a persistent way and available without the data itself \\ \midrule
\multicolumn{3}{l}{\textit{Interoperable}} \\ \midrule
I1 & 1 & RDF (with type assertions) and CSV export of comparisons\\
I2 & 2 & Reuse of ontologies where possible (ORKG core, Dublin core, RDF Data Cube Vocabulary). User responsible for other ontology reuse. \\ 
I3 & 3 & For comparisons, the compared paper metadata is linked. More references are needed and can be created by users. \\ \midrule
\multicolumn{3}{l}{\textit{Reusable}} \\ \midrule
R1 & 1 & Machine and user generated metadata is created while publishing \\
R1.1 & 1 & CC-BY SA license \\ 
R1.2 & 1 & If a registered user publishes a comparison, the user is associated with the published data \\ 
R1.3 & 2 & Users can describe contributions using domain-relevant ontologies \\ 
\bottomrule
\end{tabular}%
\end{adjustbox}
1=Yes; 2=Yes, requires user effort; 3=Partially/future work
\end{table}


\noindent
The evaluation of the FAIR principles shows that comparisons published with our service rank high in FAIRness, which can be even further increased with some effort from users. Users are mainly responsible for adding the correct information to the comparison and reuse vocabularies. Otherwise, findability, accessibility and to some extent also interoperability are largely handled by the service. 

\subsection{Performance evaluation}
In order to evaluate the performance of the overall comparison, we compared the implemented ORKG approach to a naive approach for comparing multiple resources. The naive approach compares each property against all other properties to perform the property alignment. Table~\ref{tab:sota-eval} shows the time needed to generate comparisons, for both the naive and the ORKG approach. In total, eight papers are compared with on average ten properties per paper. In the naive approach, the ``Align contribution descriptions'' step is not scaling well, since each property is compared against all others. If multiple contributions are selected, the number of property similarity checks grows exponentially. Table~\ref{tab:sota-eval} shows that the ORKG approach outperforms the naive approach. The total number of papers used for the evaluation is limited to eight because the naive approach does not scale to larger sets. 

\begin{table}[t]
\caption{Time (in seconds) to perform comparisons with 2-8 contributions using the naive and ORKG approaches.}
\begin{adjustbox}{scale=.9}
\begin{tabular}{l|lllllll}
& \multicolumn{7}{c}{\textbf{Number of compared research contributions}} \\
\hline
                  & \textbf{2} & \textbf{3} & \textbf{4} & \textbf{5} & \textbf{6} & \textbf{7} & \textbf{8} \\ \hline
\textbf{Naive} & 0.00026         & 0.1714          & 0.763          & 4.99            & 112.74          & 1772.8          & 14421           \\
\textbf{ORKG}     & 0.0035          & 0.0013          & 0.01158         & 0.02            & 0.0206          & 0.0189          & 0.0204         
\end{tabular}
\end{adjustbox}
\label{tab:sota-eval}
\end{table}

\section{Discussion \& future work}
\label{section:discussion}
One of the aims of the contribution comparison functionality is to support literature reviews and make this activity less cumbersome and time consuming for researchers. To live up to this aim, more structured contribution descriptions are needed. Existing scholarly knowledge graph initiatives focus primarily on scholarly metadata, while with ORKG we focus on making the actual research contributions machine readable. Currently, the ORKG does not yet contain sufficient contribution descriptions in order for the comparison functionally to be practically useful for researchers. Furthermore, for an evaluation of the effectiveness of certain components of the methodology (such as finding related papers or aligning similar properties), more contribution data is needed. Publishing surveys does not rely on data quantity and is therefore evaluated more extensively in this work. The performance evaluation results indicate that the comparison feature performs well. This means the technical infrastructure is in place for the literature survey service.

In the evaluation, we focused on the aspects of the system that are necessary for researchers to use the system in practice. The information representation evaluation is a straightforward evaluation to see if existing survey tables can be regenerated with the ORKG. This is a minimal requirement for researchers when using the system, since they should at least be able to recreate tables. This evaluation does not give insight to the usefulness and usability of the system, but still provides an indication that the service can be successfully used to publish literature surveys. One of the reasons for using the service is that also ``dark data'' in comparisons is published (as discussed in Section~\ref{section:motivating-example:publish-reviews}).

Another interesting aspect of the service is that published literature surveys rank high in FAIRness. Therefore, the second part of the evaluation focuses on how the FAIR principles are met. Merely publishing data as RDF is not sufficient to fully meet the FAIR principles. Hence, we conducted a more detailed evaluation that describes how the service complies with each sub-principle. Since FAIR is not a \textit{standard}, the principles are permissive and not prescriptive~\cite{Mons2017}. No technical requirements are specified. Both the implementation and evaluation of the guidelines are therefore subject to interpretation. With respect to data interoperability and reusability, certain aspects of the service can be improved. For example, to improve interoperability, the contribution data should be reusing existing vocabularies where possible. Additionally, although most of FAIRification is done by the system, the researcher is responsible for adding correct and relevant metadata while publishing a survey.

As indicated earlier, the usefulness of the presented tool depends on the number of papers present in the knowledge graph. Therefore, future work will focus on data collection, both in a crowdsourced and automated manner. We plan on extending the methodology presented in Section~\ref{section:data-collection} with automated extraction of data and tables from literature review papers. With the extracted review data, the knowledge graph can be extended more quickly than the previously presented manual method. It could form the basis of a high quality scholarly knowledge graph that contains relevant and FAIR survey table data. Furthermore, in the future we will assign (DataCite) DOIs to published surveys. They will serve as a persistent identifier for the survey data~\cite{norman2011}. 

\section{Conclusion}
Reviewing existing literature is an important but cumbersome and time consuming activity. To address this problem, we presented a methodology and service that can be used to generate literature surveys from a scholarly knowledge graph. This service can be used by researchers in order to get familiar with existing literature. Additionally, the tool can be used to publish literature surveys in a way that they largely adhere to the FAIR data principles. The presented methodology addresses multiple aspects, including finding suitable contributions, aligning contribution descriptions, visualization and publishing. The methodology is implemented within the Open Research Knowledge Graph (ORKG). Since the comparison relies on structured scholarly knowledge, we discussed how to populate the ORKG with relevant data. This is done by extracting tabular survey data from existing literature reviews. In order to evaluate whether the proposed service can be used to publish literature surveys, the original survey table representations were compared with the ones generated by our service. As the results indicate, it is possible to use the service as an addition or potentially even replacement of the current publishing approach, since the same tables can be generated. The evaluation also showed how the published literature surveys largely adhere to the FAIR data principles. This is crucial for data reusability and machine actionability. To conclude, the proposed literature comparison service addresses multiple weaknesses of the current survey publishing approach and can be used by researchers to generate,  publish and reuse literature surveys. 

\begin{acks}
This work was co-funded by the European Research Council for the project ScienceGRAPH (Grant agreement ID: 819536) and the TIB Leibniz Information Centre for Science and Technology. We want to thank Kheir Eddine Farfar for his contributions to this work.
\end{acks}

\bibliographystyle{ACM-Reference-Format}
\bibliography{refs,mendeley}

\end{document}